\documentclass[a4paper,12pt]{article}
\usepackage{amsmath,amsfonts,amssymb}

\begin{document}

\begin{center}
{\LARGE\bf The no-wall holographic model for vector
quarkonia\footnote{A talk presented at QUARKS-2016.}}
\end{center}
\bigskip
\begin{center}
{\large S. S. Afonin and I. V. Pusenkov}
\end{center}

\begin{center}
{\small Saint Petersburg State University, 7/9 Universitetskaya
nab., St.Petersburg, 199034, Russia}
\end{center}
\bigskip

\begin{abstract}
We use the no-wall holographic approach (a relative of the
soft-wall one) to construct a universal description of vector
mesons with arbitrary quark masses. The proposed model
predicts a specific dependence of the parameters of radial
Regge trajectories on the quark masses
in a reasonable agreement with the meson phenomenology and
some theoretical expectations.
\end{abstract}

\bigskip

The bottom-up holographic approach to QCD introduced in Ref.~\cite{son1}
has many interesting applications in the phenomenology of the
strong interactions. Usually this approach is applied to
the spectroscopy of light hadrons and the related physics~\cite{erlich}.
The heavy-quark sector is addressed relatively seldom
(see, e.g.,~\cite{afonin2013,schmidt} and references therein).
In this sector, only the unflavored vector mesons have a rich
established radial spectrum~\cite{pdg} that can be compared with theoretical
predictions. Our aim is a description of this spectrum
within the no-wall holographic approach~\cite{nowall}.

The action of the simplest holographic no-wall model is defined by
\begin{equation}
\label{15}
S=\int d^4\!x\,dz\sqrt{g}\left\{\sum_i\left(|DX_i|^2-m_i^2|X_i|^2\right)
-\frac{1}{4g_5^2}F_{MN}^2\right\},
\end{equation}
where
\begin{equation}
\label{16}
g=|\text{det}g_{MN}|, \qquad F_{MN}=\partial_M V_N-\partial_N V_M,
\qquad D_M X_i=\partial_M X_i - iV_M X_i,
\end{equation}
and $M,N=0,1,2,3,4$. The action~\eqref{15} is defined in the AdS$_5$
space with the metrics
\begin{equation}
\label{4}
g_{MN}dx^Mdx^N=\frac{R^2}{z^2}(\eta_{\mu\nu}dx^{\mu}dx^{\nu}-dz^2).
\end{equation}
Here $\eta_{\mu\nu}=\text{diag}(1,-1,-1,-1)$ and $R$ denotes the
AdS$_5$ radius. Below we set $R=1$ for simplicity. The holographic
coordinate $z\geq0$ has the interpretation of inverse energy
scale. The UV boundary $z=0$ represents the 4D
Minkovski space. The 5D vector field $V_M(x,z)$ is dual to the 4D
conserved vector current $j_\mu=\bar{\psi}_q\gamma_\mu\psi_q$ for
any quark flavor $q$. The gauge invariance of the action~\eqref{15}
allows to choose a convenient axial gauge $V_z=0$.
On the UV boundary, the scalar fields $X_i$ are identified
with sources of various QCD operators with canonical dimension
$i$. The corresponding 5D masses are~\cite{witten}
\begin{equation}
\label{17}
m_i^2=i(i-4).
\end{equation}
By assumption, the fields $X_i$ acquire $z$-dependent vacuum
expectation values $\langle X_i\rangle$ which represent the
$x$-independent solutions of the equation of motion,
\begin{equation}
\label{18}
\partial_z\left(\frac{1}{z^3}\partial_z X_i\right)=
\frac{m_i^2}{z^5}X_i,
\end{equation}
with the UV boundary condition
\begin{equation}
\label{19}
\left.\langle X_i\rangle\right|_{z=0}=0.
\end{equation}

The vector physical modes are defined by the plane-wave ansatz
$V_\mu(x,z)=e^{ipx}v(z)\varepsilon_\mu$. The mass spectrum of
these modes is given by the eigenvalues $M_n^2=p_n^2$ of the
corresponding equation of motion,
\begin{equation}
\label{20}
\partial_z\left(\frac{1}{z}\partial_zv_n\right)+\frac{M_n^2}{z}v_n=
\frac{2g_5^2}{z^3}v_n\sum_i\langle X_i\rangle^2.
\end{equation}
The change of variables $v_n=\sqrt{z}\,\psi_n$ brings this equation
into the Schr\"{o}dinger form
\begin{equation}
\label{21}
-\partial_z^2\psi_n+\left(\frac{3}{4z^2}+2g_5^2f(z)\right)\psi_n = M_n^2\psi_n,
\end{equation}
where
\begin{equation}
\label{22}
f(z)=\frac{1}{z^2}\sum_i\langle X_i\rangle^2
\end{equation}
determines the analogue of potential in the Schr\"{o}dinger equation.

Let us consider the dimension-two operator, $i=2$, and neglect all
others. The solution of Eq.~\eqref{18} satisfying~\eqref{19} reads
\begin{equation}
\label{23}
\langle X_2\rangle=C_2^{(1)}z^2+C_2^{(2)}z^2\ln z.
\end{equation}
Choosing $C_2^{(2)}=0$, the equation~\eqref{21} coincide
with the equation on the mass spectrum in the standard
holographic soft-wall model~\cite{son2}.

The potential in Eq.~\eqref{21} is written near the UV boundary,
where, by assumption, the holographic correspondence allows to use the QCD
inputs. In order to calculate the mass spectrum we must continue
the function $f(z)$ to the infrared domain, $z\rightarrow\infty$.
The linear radial Regge spectrum observed in the unflavored
light and heavy quarkonia~\cite{AP} can be obtained only if
\begin{equation}
\label{24}
f(z)|_{z\rightarrow\infty}=a^2z^2.
\end{equation}
The UV asymptotics of the 5D field dual to the dim2
operator provides the correct IR asymptotics automatically. 

The dependence of the linear mass spectrum
on the quark mass appears from the
v.e.v. $\langle X_3\rangle$ as the field $X_3$ is dual to the
quark bilinear operator $\bar{q}q$. The quark mass emerges from
the following AdS/CFT prescription: The solution of classical 
e.o.m. for a scalar field $\Phi$ corresponding 
to an operator $O$ of canonical dimension $i$ near the 4D 
boundary $z\rightarrow0$ has the form~\cite{kleb}
\begin{equation}
\label{25}
\Phi(x,z)\rightarrow
z^{4-i}\left[\Phi_0(x)+\mathcal{O}(z^2)\right]+z^{i}\left[\frac{\langle
O(x)\rangle}{2i-4}+\mathcal{O}(z^2)\right],
\end{equation}
where $\Phi_0(x)$ acts as a source for $O(x)$ with the v.e.v. 
$\langle O(x)\rangle$. In the real QCD, the
quark mass $m$ acts as the source for the operator $\bar{q}q$.

For the canonical dimension $i=3$, the solution of Eq.~\eqref{18}
satisfying~\eqref{19} is $\langle
X_3\rangle=C_3^{(1)}z+C_3^{(2)}z^3$. According to the
prescription~\eqref{25}, this solution can be rewritten in terms
of the physical quantities,
\begin{equation}
\label{26}
\langle X_3\rangle=\xi m z+\frac{\sigma}{2\xi}z^3,
\end{equation}
where $\sigma$ stays for the quark condensate 
$\langle\bar{q}q\rangle$ and the normalization 
factor $\xi$ is~\cite{cohen},
\begin{equation}
\label{27}
\xi^2=\frac{N_c}{4\pi^2}.
\end{equation}
The solution~\eqref{26} allows to extract explicitly the
$m$-dependent terms in~\eqref{22},
\begin{equation}
\label{28}
f(z)\rightarrow \xi^2m^2+m\sigma z^2+\tilde{f}(z),
\end{equation}
where the contribution $\frac{\sigma^2}{4\xi^2}z^6$ is absorbed
into the new sum $\tilde{f}(z)$. The Regge form of the spectrum 
appears in the case of the asymptotics
\begin{equation}
\label{29}
\left.\tilde{f}(z)\right|_{z\rightarrow\infty}=a^2z^2+\delta,
\end{equation}
which will be considered as our choice of $\tilde{f}(z)$.

We arrive then at the following equation on the mass spectrum,
\begin{equation}
\label{30}
-\partial_z^2\psi_n+\left[\frac{3}{4z^2}+2g_5^2(\sigma m+a^2)z^2+
2g_5^2\xi^2m^2+2g_5^2\delta\right]\psi_n = M_n^2\psi_n.
\end{equation}
The spectrum of Eq.~\eqref{30} is
\begin{equation}
\label{31}
M_n^2=4\sqrt2\,g_5\sqrt{\sigma m+a^2}(n+1)+2g_5^2\xi^2m^2+2g_5^2\delta.
\end{equation}
This expression demonstrates the parametric dependence of the linear
spectrum on the quark mass $m$. The spectrum~\eqref{31} can be written 
in a more compact form,
\begin{equation}
\label{32}
M_n^2=\sqrt{\alpha+\beta m}(n+1)+\gamma m^2+\delta.
\end{equation}
The value of constant $\gamma$ follows from the
relations~\eqref{27} and the normalization factor for the 5D massless
vector fields, $g_5^2=\frac{12\pi^2}{N_c}$~\cite{son1},
\begin{equation}
\label{33}
\gamma=2g_5^2\xi^2=6.
\end{equation}
The ensuing phenomenological fits are elaborated 
in Ref.~\cite{afonin2013}. It can be also shown~\cite{afonin2013}
that the behavior of the obtained slope in the chiral limit, 
$m\rightarrow0$, matches a prediction of the dual 
Veneziano-like amplitudes with the chiral symmetry 
breaking~\cite{ampl}. 

The spectrum~\eqref{32} predicts that the binding energy grows 
linearly with the quark mass in the heavy quark limit, 
$m\rightarrow\infty$, due to the relation~\eqref{33}
showing that $\gamma>4$. This behavior seems to agree
with the experimental data~\cite{pdg}. For instance, 
one has for the ground states $M_{\psi}-2m_c\approx0.56$~GeV and
$M_{\Upsilon}-2m_b\approx1.1$~GeV if $m_{c,b}$ are taken at 2~GeV
(taking $m_b$ at 5~GeV leads to $M_{\Upsilon}-2m_b\approx2.5$~GeV). 
Such a behavior admits a simple interpretation: When a
non-relativistic quark of mass $m$ is created and moves with the
velocity $v$, the binding energy must compensate its kinetic
energy $\frac{mv^2}{2}$.

It should be emphasized that the mass relation of the kind~\eqref{31} 
emerges within a whole class of holographic models and we have
presented a particular model from this class. An example of 
another model of this sort is constructed in Ref.~\cite{afonin2013}.

The proposed approach can be applied to various physical
problems including the hadron spectroscopy, finite-temperature 
and density effects, had\-ron formfactors {\it etc}. 

\underline{\it Acknowledgments}. The work was supported by the
Saint Petersburg State University research grant 11.38.189.2014
and by the RFBR grant 16-02-00348-a.

\end{document}